\documentclass[12pt]{article}

\usepackage{amssymb}
\usepackage{amsfonts}
\usepackage{amsmath}
\usepackage{txfonts}
\usepackage{hyperref}
\usepackage{lineno}
\usepackage{graphicx}

\begin{document}

\title{Bayesian Modeling of COVID-19 Positivity Rate; \\ {\it - the Indiana experience}}
\author{Ben Boukai\thanks{Correspondence: bboukai@iupui.edu}\ \  and \ Jiayue Wang\\
Department of Mathematical Sciences, IUPUI, Indianapolis,\\
IN 46202 , USA}
\date{July 7, 2020}
\maketitle

\begin{abstract}
In this short technical report we model, within the Bayesian framework, the rate of  positive tests reported by the the State of Indiana, accounting also for the substantial variability (and overdispeartion) 
in the daily count of the tests performed. The approach we take, results with a simple procedure for prediction, \textit{a posteriori},  of this rate of 'positivity' and allows for an easy and a straightforward 
adaptation by any agency tracking daily results of COVID-19 tests. The numerical results provided herein were obtained via an updatable R Markdown document.
\bigskip

\end{abstract}

%

\section{Introduction}

The \textit{Indiana State Department of Health}, (ISDH), as any other similar entity across the nation and world-wide, is closely monitoring nowadays a pandemic of the 2019 novel Coronavirus, AKA: COVID-19. According to the \textit{World Health Organizing}, (WHO), this highly contagious respiratory virus was first identified and reported in the city of Wuhan in China  in early January 2020.  Since then, this virus continues to spread and infect people around the world, including the United States. On March 11, 2020, the WHO published an assessment that COVID-19 can be characterized as a Pandemic. As of the date of this report (July 7, 2020), the \textit{John Hopkins University's Coronavirus Resource Center}, reported over 11,500,000 infected people and 538,000 deaths due to the Coronavirus, globally, of which over 2,938,000 infections and 130,000 deaths are in the USA alone. 

The state of Indiana was not spared of the contagious impact of the virus.  On March 6, ISDH confirmed the first case of COVID-19 in a Hoosier with recent travel. On March 16, the ISDH reported the first death in Indiana due to COVID-19. It subsequently worked with federal and local partners, including the \textit{Centers for Disease Control and Prevention},  (CDC), to respond to this pandemic and the grave public health situation. 
Among other responses, the ISDH created a dashboard and a data depository for tracking and reporting the daily number of COVID-19 deaths in the State as well as the daily number of tests performed and the daily number of positive cases. While there is substantial (and urgent) effort being made, world-wide, for modeling of the the death rate (or the infection fatality rate, IFR) of COVID-19, (see for example Basu (2020)), there has been very little attention given to modeling the rate of reported positive tests (often referred to rate of 'positivity'). 

In this report we model, within the Bayesian framework, the rate of  positive tests reported by the the State of Indiana, accounting also for the substantial variability of the daily number test performed. The Bayesian approach has been used successfully in other COVID-19 related studies, e.g.: Dana, at. el. (2020), or Bayes, at. el. (2020). However, to the best of our knowledge,   none of the available studies (to date) have direct relevancy to the modeling of the rate of positivity. The approach we take, provides a method for a valid prediction of this rate and allows for an easy and a straightforward adaptation by any agency tracking daily results of COVID-19 tests.

\section{The available ISDH data}

The Indiana COVID-19 data are available for retrieval through the ISDH Data Hub [10] as are reported for the state in the file \texttt{covid\_report\_date.xlsx}. It includes the daily records of the number of COVID-19 deaths cases, the total (daily) number of testings performed and the count of the positive cases, (see Appendix B for additional information).  

For the purpose of this report, we focus attention only on two of the quantities reported; the daily reported number of test performed, \textbf{COVID\_TEST} and the daily reported number of positive tests, \textbf{COVID\_COUNT} as were reported for the (successive)  $127$   daily records to date. Their ratio, the \textit{daily}  \textbf{Percentage of Positive Tests} (PPT), is used for tracking and monitoring the pandemic's progression by the state and localities. The PPT is also an important indicator of the scope and extent of the state's testing enterprise; its high value suggests that testing is conducted primarily of the sickest patients and less so of the mild or the asymptomatic cases. A lower PPT may suggest that testing has been extended to cover patients with milder or no symptoms at all. According to the  \textit{CDC Guidelines} [7], among other stipulations, a PPT $\leq 15\%$  serves as a threshold for entering Phase II of the reopening plans for the State, whereas achieving a PPT $\leq 10\%$ serves a threshold of entering Phase III of the reopening plans. 
At present, the 7-day average of this rate (of positive tests) for Indiana is  7.61\% and cumulatively, the Indiana PPT stands at 9.18\% (i.e. the total number of positive tests relative to the total number of tests performed to date, see also the ISDH Dashboard, which enabled the state to enter Phase III of the reopening plans. Thus, the importance of the appropriate modeling and tracking this percentage of reported positive tests cannot be overstated, as it has tremendous economic implications for the State and its people. In particular, constructing a valid predictive model, which predicts reasonably well the 'next-day' PPT,  would provide the State with an early sign of a looming surge in positive cases, and would  allow it to implement corrective measures and policies.  

\begin{center}{
\includegraphics[width=4in,height=3in,keepaspectratio]{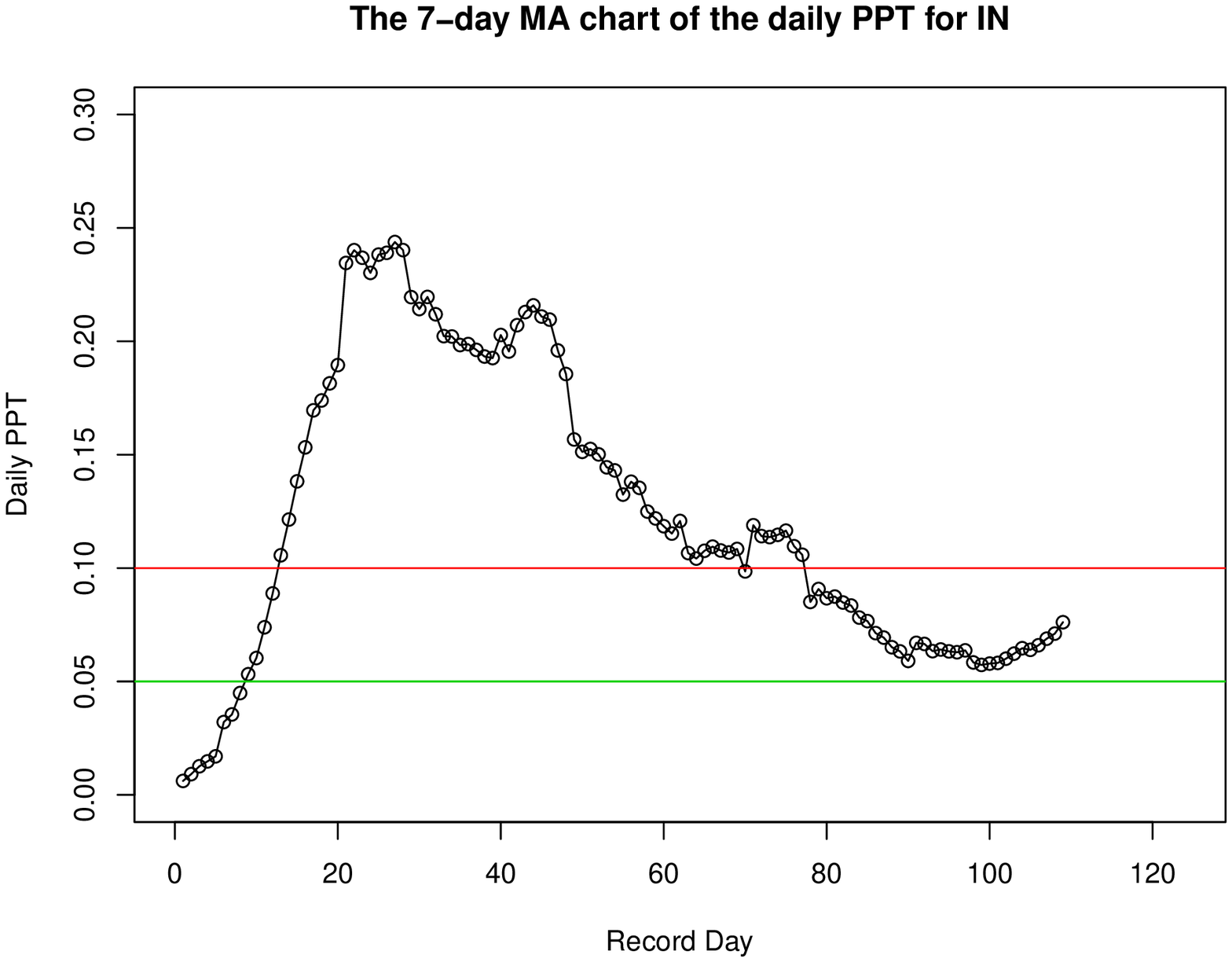}
}\end{center}
{Figure 1:}{\it The 7-day Moving Average Chart of the daily of Indiana's PPT. Also marked are the CDC's 10\% and the 5\%  Thresholds. }

\bigskip
Figure 1 above displays the time-plot of the 7-days moving average for this ratio for Indiana, starting from March 16, 2020. Since the available records for the latest few days is still updating, we included records up to the last 3 days  in the series. Thus, the data series includes $m=109$ data points out of the available $127$ records. Also marked in the figure are two horizontal lines indicating the 10\% and 5\% thresholds. In Appendix A below, we provide basic descriptive statistics concerning the daily reported number of tests and the daily reported number of positive tests as well as  of their ratio, the PPT. 

\begin{center}{
\includegraphics[width=4in,height=3in,keepaspectratio]{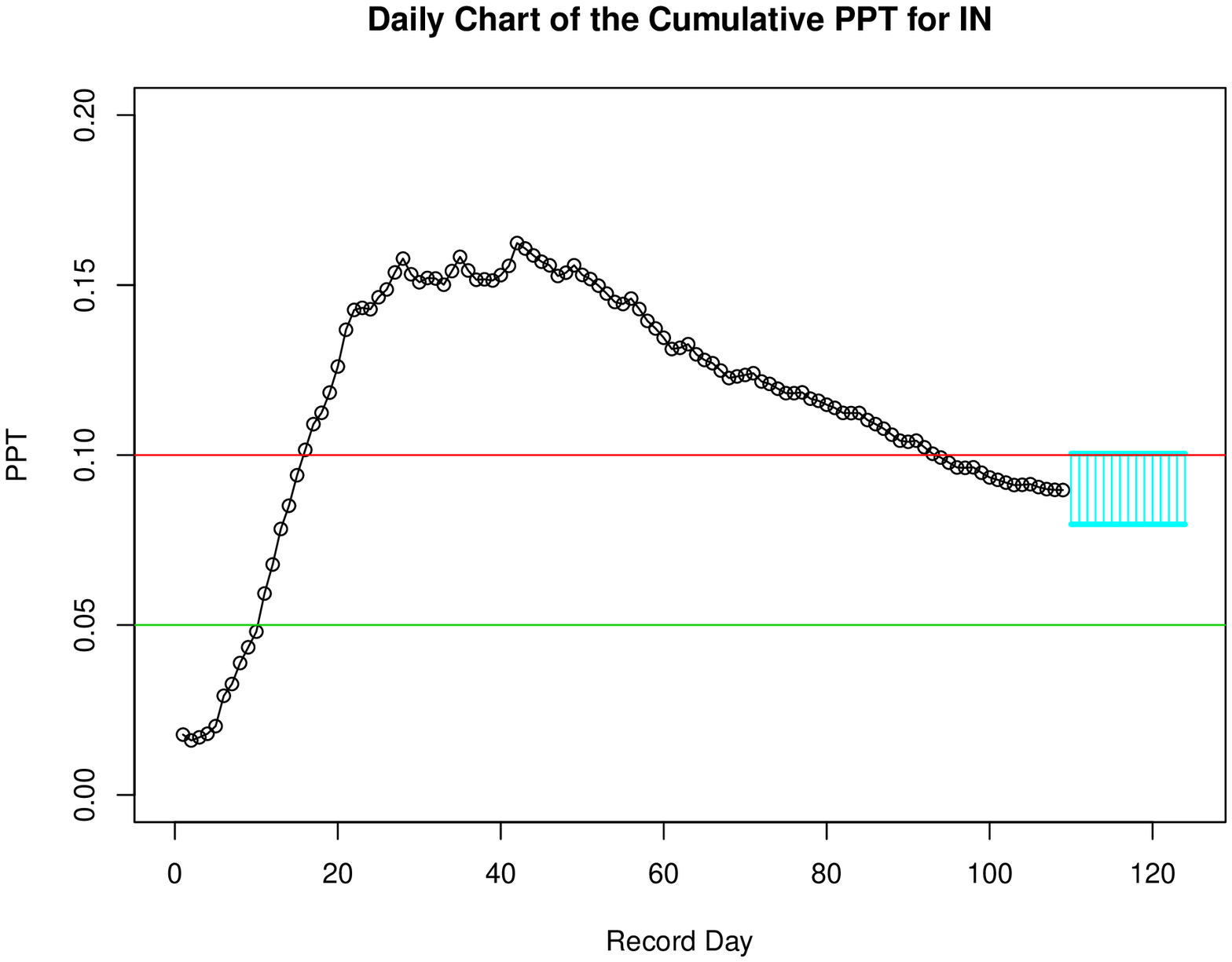}
}\end{center}
{Figure 2:}{\it The IN cumulative PPT chart. Also marked (in blue) the \textit{'next-day'} predicted (cumulative) PPT.}

\bigskip

In Figure 2 above, we present the cumulative PPT rate for Indiana over these days. We also super-imposed on the daily chart the $0.95$ (or 95\%) posterior predictive interval of $[0.07981,\ 0.1001]$ for the current Indiana (cumulative) PPT rate,  as resulted from the Bayesian predictive modeling we developed and present in this report (see Illustrations 1 and 2 below). As can be seen, the calculated posterior prediction interval contains the CDC's 10\% threshold, which could serve as an early warning flag to the State and may prompt it to initiate some mitigating measures (such as mandating masks in public) in an effort to maintain the rate of positivity below it.

\section{Bayesian modeling of Indiana's testing data}

Having retrieved the data as described above, we labeled them as $y_i\equiv${\textbf{COVID\_COUNT}  and $k_i\equiv$\textbf{COVID\_TESTS} for $i=1, \dots, m$ where $m=109$  is the total number of  observations included in the analysis. Thus, we denote the given data as $\mathcal{D}_m=\{(y_i, \, k_i), \ \ i=1, \dots, m\,\}$.

Ordinarily, when modeling the count of positive test results $y_i$ recorded out of the $k_i$ tests performed (assuming independence, risk homogeneity of the tested population and no lagging information), the binomial model would be appropriate. So that conditional on the number of tests performed, and a given $p\in (0,1)$
$$
y_i|_{k_i,\, p}\sim Binom(k_i, \, p), \ \ \ i=1, \dots, m. \eqno(1)
$$
Accordingly, given $k_1, k_2, \dots, k_m$, the daily counts of the positive tests,  $y_1, y_2, \dots, y_m$ are conditionally independent binomial random variables with $p=Pr(Test=+)$. In a similar fashion, we model the reported number of tests performed $k_1, k_2, \dots, k_m$ as (independent) Negative Binomial random  variables. That is, for given (fixed) integer  $r>0$, and $\theta\in (0,1)$,   
$$
k_i|_{\theta}\sim NegBinom(r, \, \theta), \ \ \ i=1, \dots, m.  \eqno(2)
$$
As we can see from Appendix A, the observed marginal distribution of the reported daily number of test performed is exhibiting over-dispersion features that are characteristic to  mixed-Poisson or to Negative Binomial counts. Thus, combining (1)-(2), we obtain that given $(p,\, \theta)$, the joint probability (mass) function of the $m$ pairs, $(y_i, \, k_i), \ i=1, \dots, m$, (i.e. the likelihood function),  is
$$
f(\mathcal{D}_m|p,\, \theta)=  \prod_{i=1}^m {{k_i}\choose{y_i}} p^{y_i}(1-p)^{k_i-y_i} \times {{k_i+r-1}\choose{k_i}} \theta^{k_i}(1-\theta)^{r}\\
$$
$$
\propto  \ p^{X_m}(1-p)^{N_m-X_m}\theta^{N_m}(1-\theta)^{m r}, \eqno(3)
$$
where, $\propto$ indicates proportionality of terms, up to a constant, and where $N_m=\sum_{j=1}^mk_j$  and $X_m=\sum_{j=1}^my_j$ are the cumulative reported number of tests performed and the cumulative reported positive tests. We note in passing that the cumulative PPT rate mentioned in the Introduction is merely the ratio, $\hat p_m:= X_m/N_m$.

\subsection{The conjugate Bayesian model}

The standard Bayesian predictive model in the case of Binomial-Negative Binomial counts (as in (1)-(2)), is that with the conjugate Beta-Beta joint prior distribution on $(p, \, \theta)$ (see for example Gelman et. al. (2014)). That is, in the case of the likelihood function $f(\mathcal{D}_m|p,\, \theta)$ given in (3) above, we consider the Bayesian model that assumes that for given $p$, and $N_m$, 
$$
X_m|_{N_m, \, p}\sim Binom(N_m, \, p), \qquad \text{with} \qquad p\sim Beta(a, b), \eqno(4)
$$
for some $a>0$ and $b>0$ and where, given $\theta>0$, 
$$
N_m|_{\theta}\sim NegBinom(mr,\,  \theta), \qquad \text{with} \qquad  \theta\sim Beta(c,\, d), \eqno(5)
$$
for some fixed integer $r>0$, and some $c>0$, and $d>0$. Thus, the joint prior $pdf$ of $(p, \, \theta)$ is,
$$
\pi(p, \, \theta)\propto p^{a-1}(1-p)^{b-1}\times \theta^{c-1}(1-\theta)^{d-1}\eqno(6)
$$
Accordingly, and since the joint posterior $pdf$ for $(p, \, \theta)$ **given** the data $\mathcal{D}_m$, is 
$$
\begin{aligned}
\pi(p, \, \theta|\ \mathcal{D}_m )\ \propto\  & f(\mathcal{D}_m|p,\, \lambda)\times \pi(p, \, \theta)\\
\propto\   & \pi(p\, | X_m,\ N_m) \times \pi(\theta\, | N_m), 
\end{aligned}
$$
we immediately obtain from (3)-(6) (due to the conjugacy) that **given** the data, $X_m$ and $N_m$, the (marginal) posterior distribution of $p$, denoted as $\pi(p\, | X_m,\ N_m)$,  is also a Beta distribution and the (marginal) posterior distribution of $\theta$, denoted as $\pi(\theta\, | N_m)$, is also a Beta distribution. Specifically, it follows that given $X_m$ and $N_m$, 
$$
p_{|_{X_m, N_m}}\sim Beta( a_m,\  b_m) \qquad \text{and}\qquad \theta_{|_{N_m}}\sim Beta(c_m, \, d_m) ,
$$
where these four updated parameters are given by:
$$
a_m=a+X_m, \quad b_m=b+N_m-X_m,\quad  c_m=c+N_m, \quad \text{and}\quad  d_m=d+mr. \eqno(7)
$$
Hence, the  Bayes  estimates of $\theta$ and $p$ \textbf{given} the data $(X_m, \ N_m)$, are the respective posterior means: 
$$
\hat p_{B}=E(\, p\, | X_m, \ N_m) =\frac{a_m}{a_m+b_m} \equiv \frac{a+X_m}{a+b+N_m}.
$$
and
$$
\hat \theta_{B}=E(\, \theta\, | X_m, \ N_m) =\frac{c_m}{c_m+d_m} \equiv \frac{c+X_m}{c+d+N_m+mr}.
$$
\medskip

\textbf{Remark:}  The choice in (2) for using the Negative Binomial distribution to model the reported daily number of tests $k_i$, could be seen as specific to the Indiana COVID-19 testing data, which might reflect testing capacity limitation and daily variability unique to that state. Other models that account for the observed overdispersion characteristics of the data (See Appendix A) could be used instead. For instance, one may alternatively consider the related mixed-Poisson distribution in (2).  

\medskip
\textbf{Illustration 1:}  For the Bayesian modeling of Indiana COVID-19 Testing Data, we take $a=b=1$ so that the prior mean of $p$ (the PPT) is $E(p)=0.5$ and we take $c=d=1$ so the prior mean of $\theta$ is $E(\theta)=0.5$ too.  For the given data we used $r=3$ and have, $m= 109, \ N_m= 5.22946 \times 10^5$ and $X_m=  4.6907 \times 10^4$, to calculate $\hat p_B= 0.0897$ and $\hat \theta_B=0.9993732$ as the posterior mean of $p$ and $\theta$, respectively, given the data $X_m$ and $N_m$. Clearly, the choices of values for the prior parameters $a,b,c$ and $d$, could be guided empirically (i.e. using an empirical Bayes approach), however, in light of the magnitude of  the given data counts, $X_m$ and $N_m$, this model exhibits very little sensitivity to the choices of these prior parameters.

\section{The posterior predictive model}

Having observed $(X_m, \, N_m)$, the \textit{posterior predictive distribution} ,  under the Bayesian model (3)-(6) and given $(X_m,\ N_m)$,  of a `new' (or `future') observation on the number of positive cases, $Y^*$,   out of a given $K^*=k^*$ new tests is the the beta-binomial distribution, 
$$
Pr(Y^*=y^*|K^*=k^*, X_m,\, N_m)=   \int_{0}^1P(y^*=y^*|p, \, K^*=k^*)\times \pi(p\, | X_m, \ N_m)dp 
$$
$$
={{k^*}\choose{y^*}}\times \frac{B(y^*+a_m,\  b_m+k^*-y^*)}{B(a_m,\ b_m)}, \eqno(7)
$$
for $y^*=0, 1, \dots, k^*$. We denote this (predictive) distribution for $Y^*$, given $K^*=k^*$, as
$$
Y^*\sim BetaBinom(k^*, a_m, \ b_m),
$$
with $a_m$ and $b_m$ as in (7). The corresponding posterior predictive mean and variance of $Y^*$ are given by 
$$
 E(Y^*|K^*=k^*, X_m, N_m)= k^*\times \frac{a_m}{a_m+b_m}\equiv k^*\times \hat p_{B},  
$$
and
$$
Var(Y^*|K^*=k^*, X_m, N_m) =  k^*\times \hat p_{B}(1-\hat p_{B})\times \frac{a_m+b_m+k^*}{a_m+b_m+1},
$$
respectively. 

In a similar manner we obtain the posterior predictive distribution  under this Bayesian model and given $(X_m,\ N_m)$,  of a `new' (or `future') number of tests $K^*$ is the Beta-Negative Binomial distribution. In fact, with $K^*|_\theta\sim NegBinom(r, \theta)$, we have
$$
Pr(K^*=k^*|\, X_m,\, N_m)=   \int_{0}^1P(K^*=k^*|\, \theta)\times \pi(\, \theta, | X_m, \ N_m)d\theta 
$$
$$
=  {{k^*+r-1}\choose{k^*}}\times \frac{B(k^*+c_m,\  r+d_m)}{B(c_m,\ d_m)}.\eqno(8)
$$
for $k^*=0, 1, 2, \dots$.  We denote this (posterior predictive) distribution for $K^*$ as
$$
K^*\sim BetaNegBinom(r, c_m, \, d_m),
$$
with $c_m$ and $d_m$ as in (7). The corresponding posterior predictive mean and variance of $K^*$ are given by 
$$
 \mu^*:= E(K^*|X_m,\,  N_m)={ r\,  c_m\over d_m-1}\equiv r \times {c+N_m\over d-1+mr},  
$$
and
$$
Var(K^*|X_m,\, N_m) =  \mu^* \times {(d_m+r-1)(c_m+d_m-1) \over  (d_m-1)(d_m-2)}.
$$
 
It is straightforward to see that the joint posterior predictive probability (mass) function of $(Y^*,\ K^*)$, can easily be obtained from expressions (7) and (8) is 
$$
\begin{aligned}
  Pr(Y^*=&  y^*, \ K^*= k^*\, |X_m,\, N_m)= \\
= & Pr(Y^*=y^*|K^*=k^*, X_m,\, N_m)\times  Pr(K^*=k^*|\, X_m,\, N_m)=\\
= & {{k^*}\choose{y^*}}\times \frac{B(y^*+a_m,\  b_m+k^*-y^*)}{B(a_m,\ b_m)}\times  {{k^*+r-1}\choose{k^*}}\times \frac{B(k^*+c_m,\  r+d_m)}{B(c_m,\ d_m)} \quad (9)
\end{aligned}
$$

Clearly, along with the  value of $Y^*$ as the number of  positive tests predicted out of  the  predicted number of tests, $K^*$, one may obtain their ratio, $p^*:= Y^*/K^*$, as the rate of positive tests predicted next, given the data. While an explicit expression for the posterior predictive distribution of $p^*$ is not readily available, it may be estimated, quite accurately, through Monte-Carlo simulations. Towards that end, we denote by $Q_m^*(\cdot)$ the posterior predictive $cdf$ of $p^*$, given the data $\mathcal{D}_m$. That is, for any $t\in \mathbb{R}$, 
$$
Q_m^*(t):= Pr(\, p^*\leq t\, | X_m, \, N_m), 
$$
and recall that the $\alpha^{th}$ percentile ($\alpha\in (0,1)$), of this distribution, is defined as
$$
t^*_\alpha:= \inf\{\, t,\, \ \ \text{s.t.} \ \  Q^*_m(t)\geq \alpha\, \}.  \eqno(10)
$$
Thus, when available, the interval, $[t^*_{1-\alpha}, \  t^*_\alpha]$ serves as a $1-2\alpha$ posterior prediction interval for $p^*$ given the data $\mathcal{D}_m$, so that
$$
Pr(t^*_{1-\alpha} \leq p^* \leq  \  t^*_\alpha\, |\, X_m, \, N_m)=1-2\alpha.
$$

\subsection{Estimating the predictive distribution of the PPT}

As was mentioned in the previous section, while explicit expression for $Q_m^*(\cdot)$, the posterior predictive $cdf$ of $p^*$, given the data $\mathcal{D}_m$, is not available, it may be estimated via Monte-Carlo simulations which exploit   the explicit expression, in (9), for the joint posterior predictive distribution of $(Y^*, \, K^*)$. Given the data, $\mathcal{D}_m$ and with parameters $a_m, b_m, c_m$ and $d_m$ and with $r$ as above,  generate a random sample of a large size $B$ ($B=5,000$, say) from $Q_m^*(\cdot)$ as follows:  

\begin{itemize}
\item[1)] Given $N_m$, draw $K^*\sim BetaNegBinom(r, \, c_m, \, d_m)$;
\item[2)] Given $X_m, N_m$ and $K^*=k^*$,  draw $Y^*\sim BetaBinom(k^*, a_m, \ b_m)$ to obtain the pair $(y^*, \, k^*)$;
\item[3)] Calculate the simulated predicted PPT as $p^*= y^*/k^*$.  
\item[4)] Repeat steps (1)-(3) $B$ times so as to form $p^*_1, p^*_2, \dots , p^*_{{_B}}$ as a random sample from $Q_m^*(\cdot)$.  
\end{itemize}

Having obtained the random sample $p^*_1, p^*_2, \dots , p^*_{{_B}}$, we estimate $Q_m^*(\cdot)$ by its empirical version 
$$
\hat Q_{B,m}^*(t)={1\over B}\sum_{i=1}^BI[p_i^*\leq t], \qquad \forall t\in \mathbb{R}.
$$
Accordingly, and in similarity to (10), we estimate the $\alpha^{st}$ percentile of $Q_m^*(\cdot)$ by
$$
\hat t^*_\alpha:= \inf\{\, t,\, \ \ \text{s.t.} \ \  \hat Q^*_{B, m}(t)\geq \alpha\, \}.
$$

A simple R script (see Appendix C) which utilizes the built-in functions \texttt{rbnbinom()} (for the $BetaNegBinom(\cdot)$ distribution) and \texttt{rbbinom()} (for the $BetaBinom(\cdot)$ distribution), of the \texttt{extraDistr} package produces the simulated sample from the (respective) posterior predictive distribution of $(Y^*,\, K^*)$ and the corresponding predicted PPT, $p^*=Y^*/K^*$.

\medskip

\textbf{Illustration 2:}  We continue with the same prior parameterization used in Illustration 1, of  $a=b=1$, $c=d=1$ and $r=3$. Recall that the given data $\mathcal{D}_m$, yields,  $m= 109, \ N_m= 5.22946 \times 10^5$ and $X_m=  4.6907 \times 10^4$.  We first simulated $B=5000$ sample values for the respective predictive distributions of $(Y^*,\,  K^*)$ and of $p^*$ and used these simulated values to estimate the posterior predictive distribution of $p^*$ , which in turn, was used it to obtain the 95\% posterior prediction interval, $[0.07981,\ 0.1001]$ for the `next-day' Indiana's PPT as was displayed in Figure 2. The  means and standard deviations of the (estimated) posterior predictive distributions of $Y^*, K^*$ and $p^*$, along with the corresponding 95\% prediction interval are presented in Table 1 below.

\begin{center} 
\begin{tabular}{ccccc}
	  & & &   Prediction  & Bounds  \\ \hline\hline
	  &   Mean       & SD           &  2.50\%    &  97.50\%  \  \\ \hline
\hline
$Y^*$	  &	432.460  &	250.518	&  87.0      & 1054.0  \  \\ \hline
$K^*$	  &	4819.909 &	2782.975&  11666.4   & 986.9    \\ \hline
$p^*$	  &	0.090    &	0.005	&  0.07981 & 0.1001   \  \\ \hline\hline
\end{tabular}
\medskip

{Table 1:} {\it Summarizing the posterior predictive distributions and the corresponding 95\% prediction interval   }
\end{center}

\begin{center}
  \includegraphics[width=4in,height=3.5in,keepaspectratio]{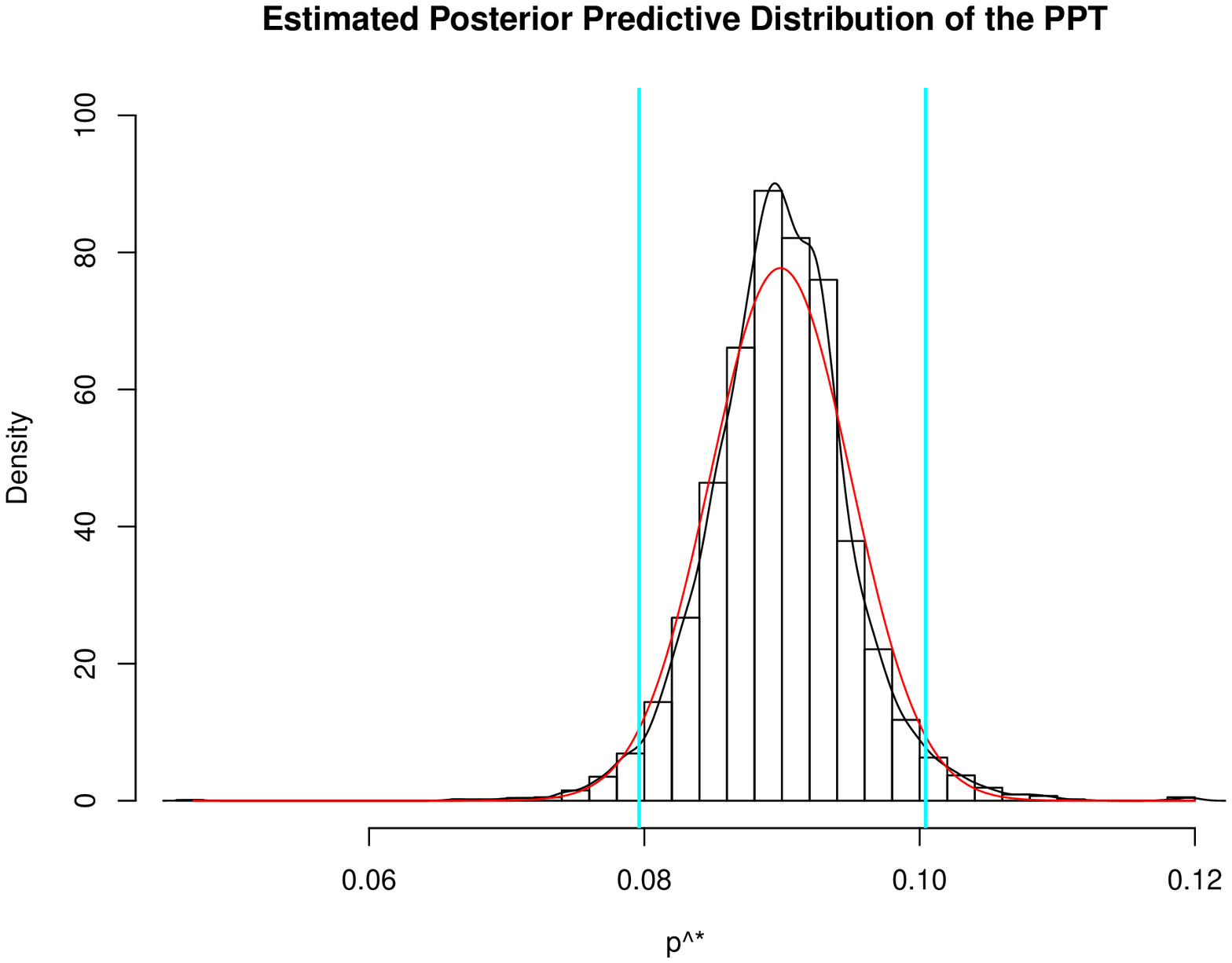}
\end{center}
{Figure 3:} {\it The estimated (Monte-Carlo) posterior predictive distribution of Indiana PPT, $p^*$, along with the marked (in blue) the 95\% prediction interval}

\bigskip

Figure 3 above, displays the Monte-Carlo histogram of that predictive distribution, along with a nonparametric and a normal density (in red) approximations.  Also marked are the corresponding bound for the $95$\% posterior prediction interval for $p^*$. 
The Monte-Carlo marginal (posterior) distribution of $K^*$ is displayed in Figure 4 and that of $Y^*$ in Figure 5. 

\begin{center}
  \includegraphics[width=4in,height=3.5in,keepaspectratio]{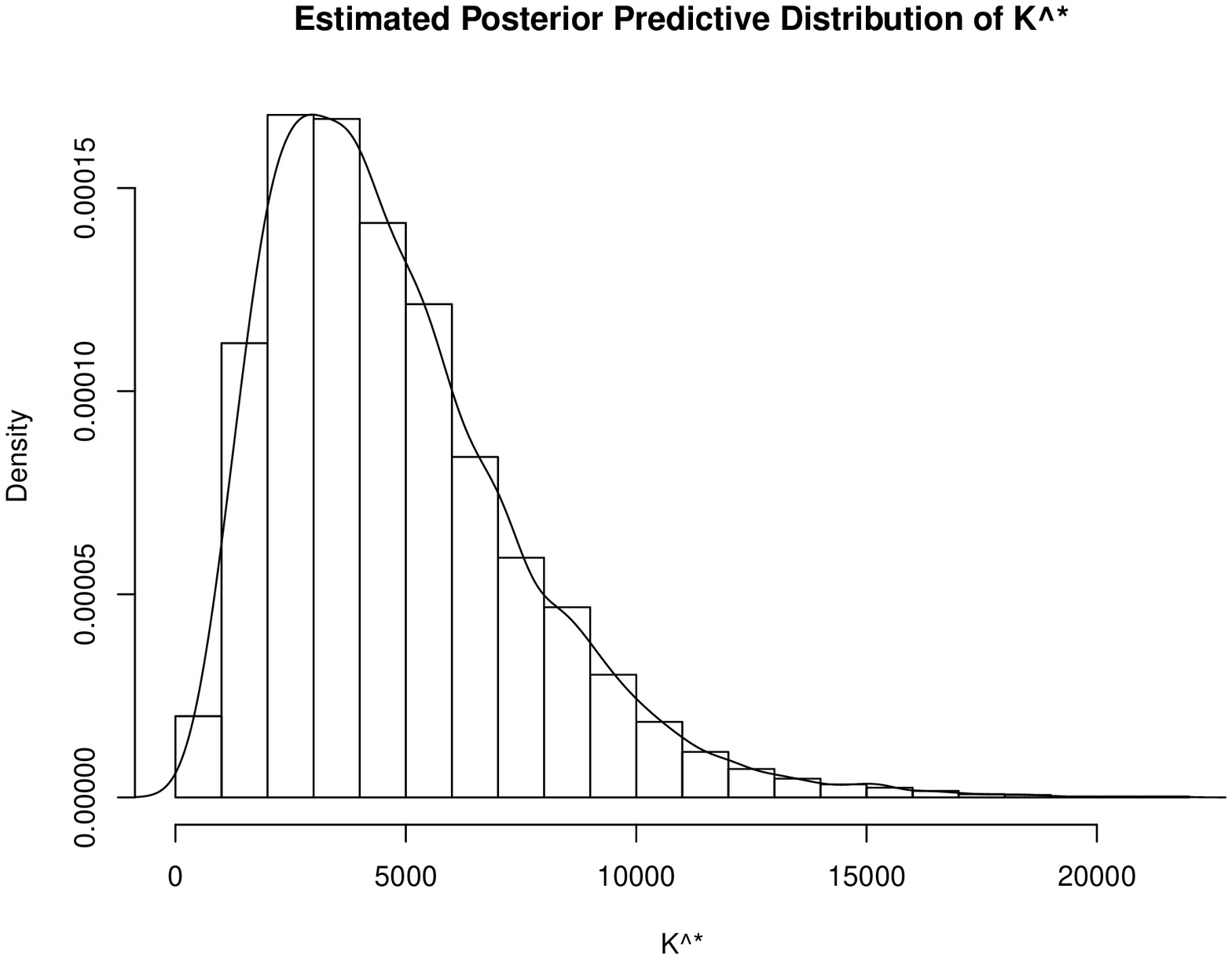}
\end{center}
{Figure 4:} {\it The estimated (Monte-Carlo) posterior predictive distribution of $K^*$, along with its none-parametric density estimate conforming with the Negative Binomial distribution.}

\begin{center}
  \includegraphics[width=3.5in,height=3in,keepaspectratio]{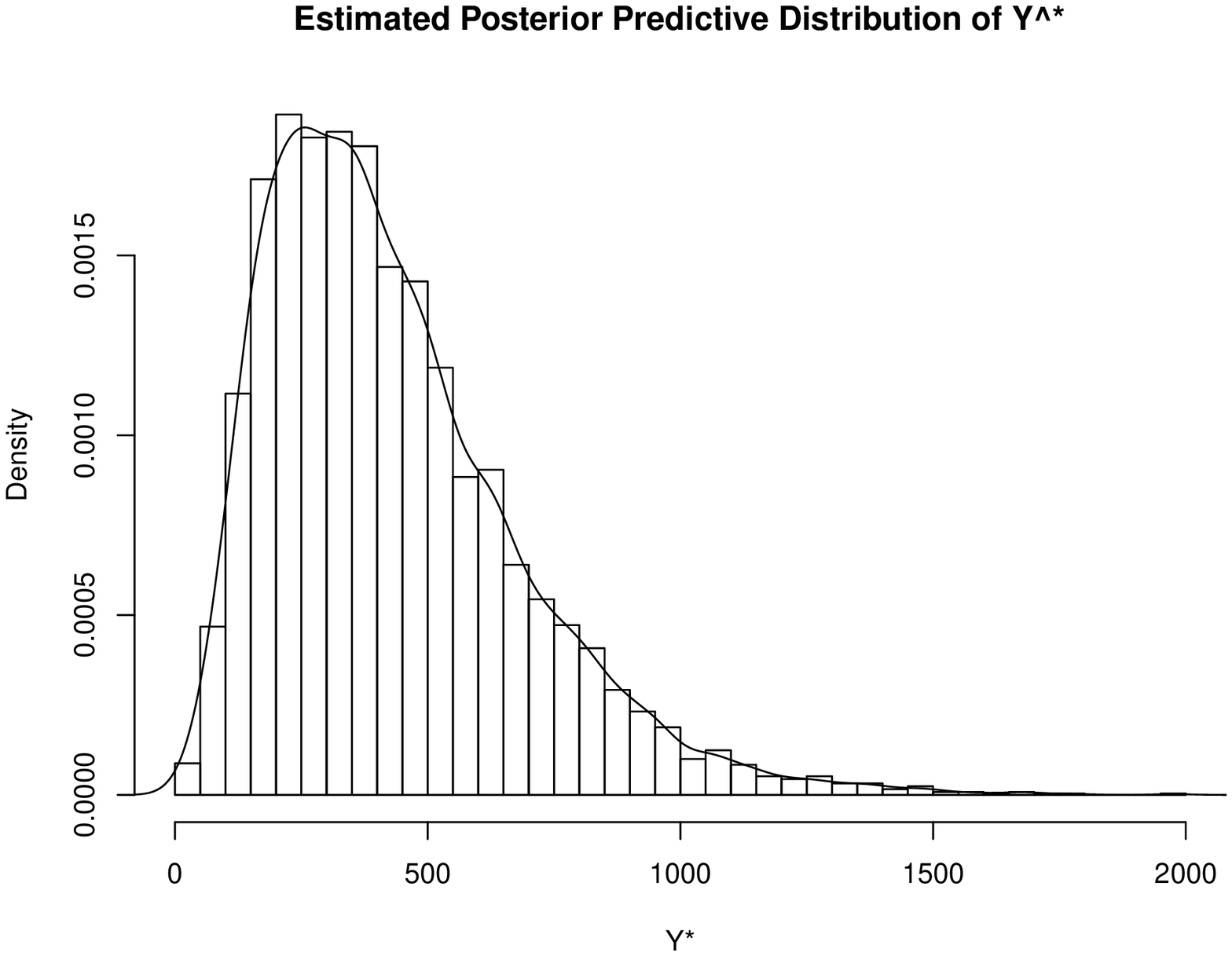}
\end{center}
{Figure 5:} {\it The estimated (Monte-Carlo) posterior predictive distribution of $Y^*$, along with its none-parametric density estimate conforming with the Negative Binomial distribution.}
\bigskip

We conclude this illustration with Figure 6, where we display the posterior prediction intervals for the PPT as were calculated for each report day in the series. That is, based on the given data on the $n^{th}$ day, $\mathcal{D}_n$, we calculated the 95\% posterior predicted  interval for the PPT on the $(n+1)^{th}$, day (the "next" day), for each $n=2, 3, \dots, m$ of the $m=109$ days available in the data set.  As can be seen, each of the daily calculated PPT (as in Figure 2), fell well within the corresponding prediction interval (marked in red in Figure 6) as was calculated based on the previous' days data, thus providing also a partial validation for the applicability of this Bayesian approach (with its underlying assumptions) to these COVID-19 count data.

\begin{center}
  \includegraphics[width=3.5in,height=3in,keepaspectratio]{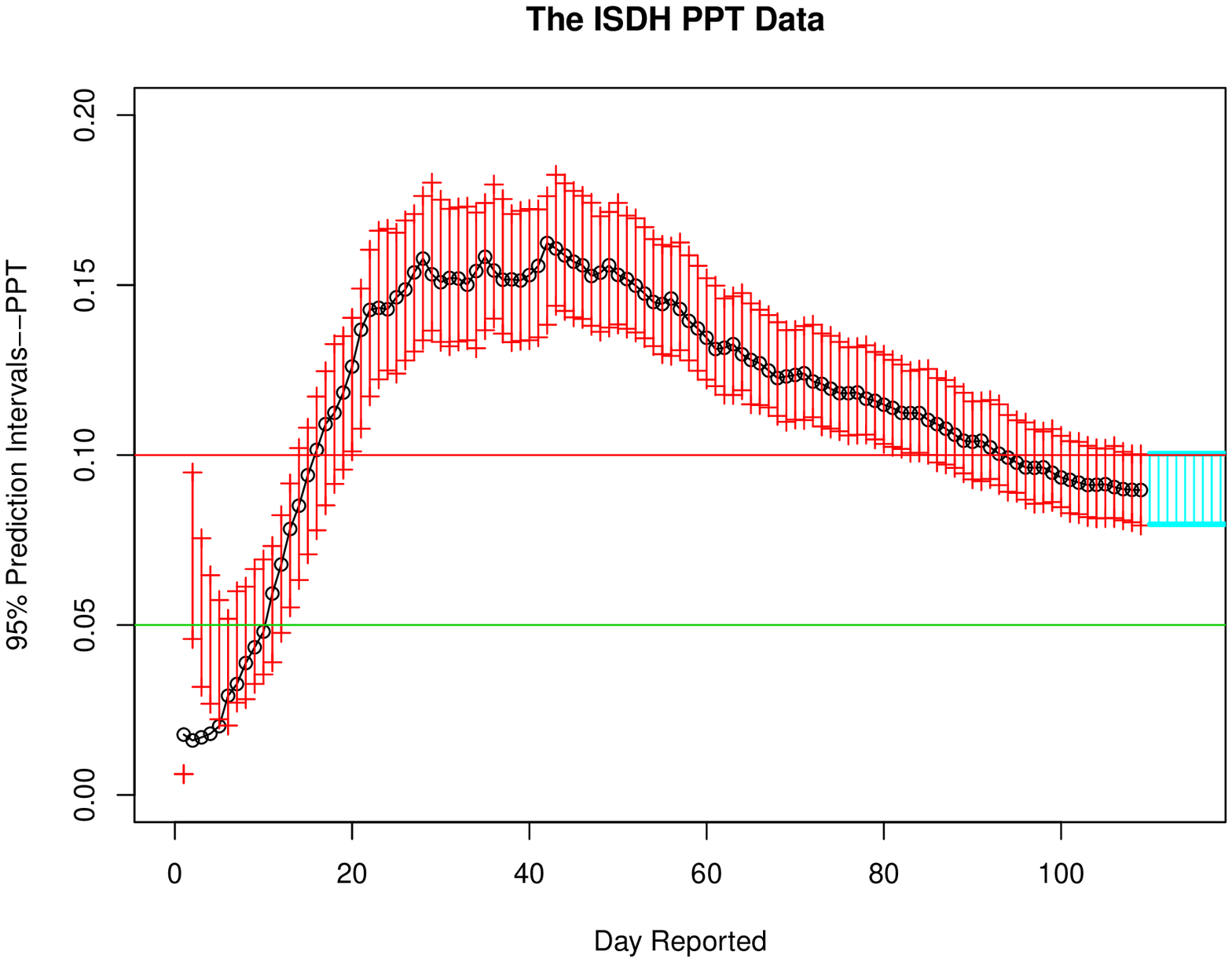}
\end{center}
{Figure 6:} {\it The IN cumulative daily PPT chart, along with its daily posterior predictive interval for each of the recent 109 days.}

\bigskip

\section{Appendices}
\subsection{Appendix A}
We provide here the basic descriptive statistics for the daily reported number of test performed, \textbf{COVID\_TEST}, the daily reported number of positive tests, \textbf{COVID\_COUNT}  and the daily \textbf{PPT} (their ratio) for the  (successive) $109$ daily records since March 16, 2020 and the most recent 3 days.  
{\small
\begin{center} {
\begin{tabular}{ccccc}
\hline\hline
&        & COVID\_TEST& COVID\_COUNT&   PPT\\ 
\hline\hline
&Min.    &    652.000&       4.000& 0.006\\
&1st Qu. &   2033.250&     348.000& 0.057\\
&Median  &   3907.000&     434.000& 0.097\\
&Mean    &   4788.564&     430.709& 0.125\\
&3rd Qu. &   7159.500&     550.000& 0.154\\
&Max.    &  11391.000&     946.000& 0.511\\
&SD  & 2999.324 & 173.064 & 0.098\\
\end{tabular}
\medskip

{Table 2:} {\it Summary Statistics of Indiana COVID-19 testing data}
}\end{center}
}

\begin{center}
  \includegraphics[width=3.5in,height=3in,keepaspectratio]{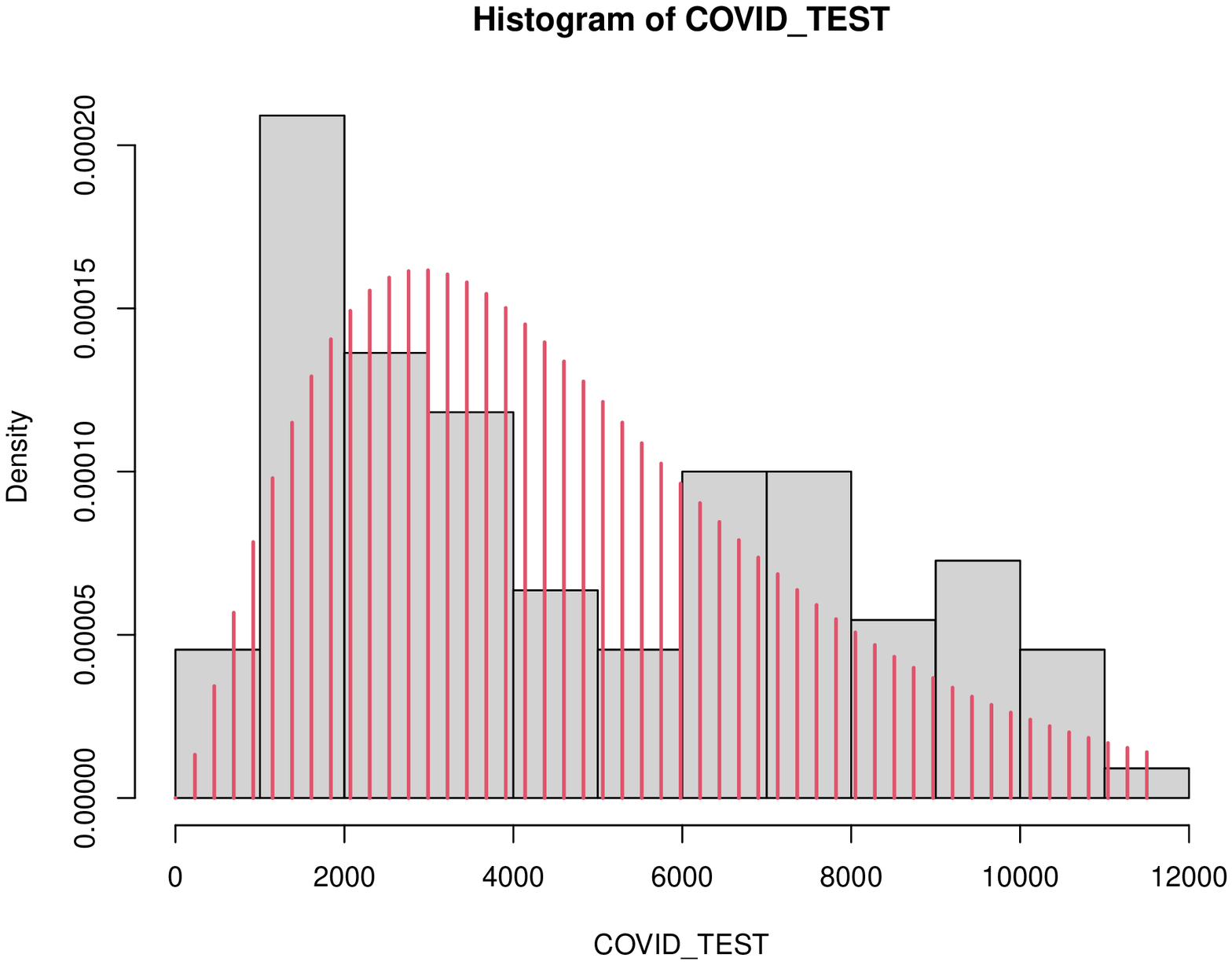}
\end{center}
{Figure 7:} {\it The histogram of \textbf{COVID\_TEST} along with the 'fitted' Negative Binomial probabilities.}

\bigskip

\subsection{Appendix B}
The Indiana Covid-19 data are available for retrieval through the  \textit{ISDH data hub}[10] as are reported for the state in the file \texttt{covid\_report\_date.xlsx}. It includes the daily records (as columns) on:
 \small{
\begin{itemize}
  \item DATE: Date of the event, it is equal to the investigation starting date for positive cases; the date of death for deaths; the coalesce of specimen date and report date for testings (if specimen collected date is unknown, use report date), respectively
  \item COVID\_TEST: Total number of testings (i.e. number of new people tested on the date. Indiana residents only)
\item DAILY\_DELTA\_TESTS: 	The number of most recent (i.e latest report) new testings that are reported into the testing pool. The date of specimen collected is typically earlier than the report date
\item DAILY\_BASE\_TESTS:	The number of tests from the last report. Records might be removed due to information correctness
\item COVID\_COUNT:	Total number of positive cases (i.e. number of patients who started investigation for their positive report) on the date
 \item DAILY\_DELTA\_CASES:	The number of most recent(i.e latest report) new positive cases that are reported into the positive case pool. The investigation starting date could be earlier than the report date due to necessary process
 \item DAILY\_BASE\_CASES:	The number of positive cases from the last report. Records might be removed due to information correctness
  \item COVID\_DEATHS	Total: number of deaths on the date
  \item DAILY\_DELTA\_DEATHS	The number of most recent (i.e latest report) new death cases that are reported into the death case pool. The date of death could be earlier than the report date due to necessary process and confirmation
 \item DAILY\_BASE\_DEATHS:	The number of deaths from the last report. Records might be removed due to information correctness
  \item COVID\_COUNT\_CUMSUM:	The cumulative number of positive cases as of the report date
  \item COVID\_DEATHS\_CUMSUM:	The cumulative number of deaths as of the report date
  \item COVID\_TEST\_CUMSUM:	The cumulative number of tests as of the report date
\end{itemize}
}  

\subsection{Appendix C}
A simple R script to obtain the Monte-Carlo sample from $Q_m(\cdot)$, 
\vskip -50pt
\begin{verbatim}
library(extraDistr)
SimPPT<-function(m, Xm, Nm, am, bm, cm, dm, r0, BB=5000){
Kstar<-rbnbinom(BB, r0, dm, cm)
Ystar<-NULL
for(j in 1:BB){
Ystar[j]<-rbbinom(1, Kstar[j], am, bm)}
Pstar<-Ystar/Kstar
out<-data.frame(Ystar, Kstar, Pstar)
return(out)}
\end{verbatim}


\end{document}